\documentclass[prx, twocolumn, aps, reprint, floatfix,
 amsmath,amssymb,
 aps,
]{revtex4-2}

\usepackage{graphicx}
\usepackage{dcolumn}
\usepackage{bm}
\usepackage{hyperref}
\usepackage{comment} 
\usepackage{color}

\begin{document}

\preprint{APS/123-QED}

\title{
Spin and Density Modes in a Binary Fluid of Light}

\author{Clara Piekarski}
 \email{clara.piekarski@lkb.upmc.fr}
\author{Nicolas Cherroret}
\email{nicolas.cherroret@lkb.upmc.fr}
\author{Tangui Aladjidi}%
\author{Quentin Glorieux}%
 \email{quentin.glorieux@sorbonne-universite.fr}
\affiliation{Laboratoire Kastler Brossel, Sorbonne Universit\'e, CNRS,
ENS-PSL Research University, Coll\`ege de France, 4 Place Jussieu, 75005 Paris, France}

\begin{abstract}

We present the experimental observation of spin and density modes in a binary mixture of superfluids of light. 
A miscible Bose-Bose mixture with repulsive interactions is obtained by propagating, in the paraxial limit, the two circular polarization components of a laser through a non-linear hot atomic vapor. 
Controlling the intensity and phase for both polarizations allows us to selectively excite the fundamental modes of the mixture.
Using a Bragg-like spectroscopy technique, we measure the dispersion relation and identify two distinct branches with different speeds of sound corresponding to the spin and density modes.
At large photon density, we observe a crossing of these branches, which is due to an effective photon-photon interaction beyond the two-body regime and related to the saturation of the medium nonlinearity.
This novel degree of freedom allows for precise control over the ratio of spin and density sound velocities and provide new insights into the control of binary mixtures collective dynamics.

\end{abstract}

\maketitle

\section{\label{sec:level1}Introduction}

Collective excitations are a key concept in many-body systems, capturing emergent macroscopic behaviors arising from microscopic particle interactions.
In superconductivity, for example, phonons mediate electron pairing, leading to the formation of a macroscopic quantum state with zero electrical resistance \cite{Tinkham2004}.
In superfluidity, excitations like phonons or rotons underpin the frictionless flow of liquid helium \cite{Pines1996}. 
Similarly, collective Bogoliubov excitations govern the behavior of dilute ultra-cold atomic gases \cite{Pitaevskii2016}. 
Since the celebrated measurements of their dispersion in the early 2000s in Bose gases \cite{Steinhauer2002, Vogels2002}, these excitations have been identified as the cornerstone of many phenomena in quantum fluids, such as superfluid flow through obstacles \cite{Desbuquois2006}, quasi-long-range coherence \cite{Hadzibabic2006, Clade2009, Tung2010} and thermalization \cite{Gring2012, Mallayya2019,Regemortel2018, Duval2023}. 

In recent years, a new frontier in the collective dynamics of quantum gases has emerged with the realization of binary superfluid mixtures, in which the strength of the inter-species interaction $g_{12}$ can be tuned independently of the intra-species interaction $g$ \cite{Baroni2024}.
When the intra-component interaction $g$ is repulsive ($g>0$) and the inter-component interaction $g_{12}$ is attractive ($g_{12}<0$), these systems reveal unique phenomena, such as self-bound quantum droplets \cite{Semeghini2018, Cabrera2018}, or beyond mean-field corrections in the equations of state \cite{Lavoine2021}. 
In binary mixtures where all interactions are repulsive ($g$ and $g_{12}$ both positive), the relative magnitude of these interactions determines whether the system is in a miscible state ($g>g_{12}$) or a non-miscible state ($g<g_{12}$)  \cite{Papp2008, Baroni2024}. 
Compared to the single-component gas, the mixing dynamics gives rise to two distinct collective excitations: the density mode, corresponding to perturbations in the total gas density, and the spin mode, corresponding to perturbations in the difference of the density of the two components. Each mode is characterized by a distinct speed of sound \cite{blakie1999dressed,tommasini2003bogoliubov,berman2001manipulating,abad2013study}. 

Quantum fluids of light provide a rich perspective for exploring quantum gases and collective excitations using the tools of non-linear and quantum optics \cite{carusotto2014superfluid,glorieux2023hot}.
In this approach, the propagation of light through a $\smash{\chi^{(3)}}$ material is described by the non-linear Schrödinger equation (NLSE), which can be mapped onto the Gross-Pitaevskii equation which describes the quantum dynamics of ultra-cold Bose gases. 
This mapping has been verified by optical measurements of the Bogoliubov spectrum \cite{fontaine2018,fontaine2020interferences,vocke2015experimental}, which paved the way for the study of a variety of fluid-like phenomena such as superfluidity of light \cite{fontaine2018,michel2018superfluid,wimmer2021superfluidity,ferreira2018superfluidity}, thermalization and dynamic phase transitions \cite{Neven2018, Larre2018, Bardon-brun2020, abuzarli2022nonequilibrium, Cherroret2024}, shock waves \cite{abuzarli2021blast,bienaime_quantitative_2021}, and vortex dynamics \cite{azam2022vortex,baker2023turbulent,ferreira2024exploring}. 
Recent theoretical works \cite{Martone2021, Martone2023} have investigated the dynamics of two-component fluids of light, but to date binary mixtures of photon fluids have not been experimentally realized.

Here, we present the experimental realization of a Bose mixture of two miscible interacting superfluids of light, using the two circular polarizations $\sigma^+$ and $\sigma^-$ of the electric field propagating along $z$ in a non-linear hot atomic vapor.
The paraxial approximation gives a mass to the photons in the transverse $(x,y)$ plane and effective photon-photon interactions arise from the light-matter interaction \cite{carusotto2014superfluid,fontaine2018}.
Although the observation of spin and density modes has proven challenging in atomic mixtures \cite{kim2020observation,cominotti2022observation}, in this work we show how to selectively or simultaneously excite the eigenmodes of the mixture through optical control of the density and phase of the two species.
We demonstrate a decoupling between the spin and density modes and, using a Bragg-like spectroscopy technique \cite{piekarski2021measurement}, we measure the dispersion relations for the spin and density modes and observe two distinct sound velocities. 

In atomic Bose mixtures with positive $g$ and $g_{12}$, the speed of sound of the density mode exceeds the speed of sound of the spin mode.
A striking property of the photonic mixture is that we observe the spin branch passing \emph{above} the density branch when operating at a high photon density, although we measure $g_{12}$ to be intrinsically positive. 
We show theoretically that this phenomenon arises from effects beyond the two-body contact interaction that are associated with the saturation of the optical non-linearity in the vapor.
These effects have no direct counterparts in atomic gases. 
Through a controlled increase of the non-linearity, we experimentally explore the spectrum of collective excitations of the mixture, from the low-intensity regime, in which two-body contact interactions dominate, to the high-intensity regime, where the spin and density branches are inverted. 
The tunable control of the sound velocities opens up new opportunities for manipulating collective excitations and studying different regimes of interactions in two-component quantum fluids.\\

The article is organized as follows:  in Sec. ~\ref{Sec:fluidmix_exp}, the general method for observing a binary mixture of fluids of light in a non-linear vapor is introduced.
We present experimental measurements of the intra- and inter-species interaction strengths, $g$ and $g_{12}$. 
We then show how to selectively excite the spin and density modes of the mixture and we experimentally extract their dispersions using a Bragg-like spectroscopy technique.
In Sec.~\ref{sec:theory},  we develop a Bogoliubov theory for fluid-of-light mixtures that accounts for effects beyond the two-body interaction by incorporating the saturation of the non-linearity.
We then present the experimental validation of the predicted crossing of the spin and density branches at high intensities.
Finally, in Sec.~\ref{Sec:summary} we summarize our findings and provide future directions for this work. 

\section{realizing an optical fluid mixture }
\label{Sec:fluidmix_exp}

\subsection{Polarized fluids of light}
In the paraxial regime, it is established that the evolution of the electric field envelope is governed by a non-linear Schr\"odinger equation (NLSE), where the Kerr non-linearity of the vapor induces effective photon-photon interactions.
This approach has been used for paraxial fluids of light to study 2D quantum gases with the propagation direction $z$ playing the role of time \cite{carusotto2014superfluid}.
However, in a fluid of light, the description in terms of a \emph{single} NLSE is only valid when the laser field is linearly or circularly polarized. 
For the more general case of an elliptically polarized field, the tensor nature of the atomic susceptibility tensor must be taken into account. 
This leads to a system of \emph{two} coupled NLSEs for the two circular components of polarization, $E_+$ and $E_-$ \cite{BOYD2008207}:
\begin{align}
    i  \partial_z E_\pm &= \left( -\frac{\nabla_\perp^2}{2k_0}+g |E_\pm|^2+g_{12} |E_\mp|^2 \right) E_\pm,
    \label{eq:cnlse} 
    \end{align}
where $k_0$ is the wavevector and $\nabla_\perp$  is the gradient operator in the transverse plane $(x,y)$.
The system (\ref{eq:cnlse}) is fully analogous to the coupled Gross-Pitaevskii equations that describe symmetric binary mixtures of atomic Bose fluids \cite{Pitaevskii2016}, where the two fluids are represented here by the two circular polarization components of light, and $k_0$ acts as an effective particle mass.
In the optical system, $g=-k_0 \epsilon_0 c{n_2}/2{n_0}$ and $g_{12} = -k_0 \epsilon_0 c{n_{12}}/2{n_0}$ can be viewed as effective photon-photon interaction parameters, which respectively describe the intra- and inter-component interactions of the two fluid components. 
They depend on the linear refractive index $n_0$  of the vapor, and on two non-linear index coefficients, $n_2=6\chi_\text{1122}/\epsilon_0 c$ and $n_{12}=6(\chi_\text{1122}+\chi_\text{1221})/\epsilon_0 c$, which are related to the two independent coefficients $\chi_\text{1122}$ and $\chi_\text{1221}$ of the third-order susceptibility tensor characterizing the (isotropic) non-linear vapor \cite{BOYD2008207}.

The single-component fluid of light scenario is recovered in two configurations: 
In the first one, exploited in previous experiments \cite{fontaine2018,fontaine2020interferences,abuzarli2022nonequilibrium,baker2023turbulent}, one uses a linearly-polarized beam, with, e.g., $E_+=E_-=E/\sqrt{2}$. 
In the second one, a circularly-polarized beam is used, with, e.g., $E_+=E$ and $E_-=0$. 
In both cases, the coupled system (\ref{eq:cnlse}) reduces to the single NLSE: 
\begin{align}
    i  \partial_z E = \left( -\frac{\nabla_\perp^2}{2k_0}  +g_\text{circ,lin} |E|^2 \right) E,
\end{align}
where $g_\text{circ}=g$ for circularly-polarized light, and $g_\text{lin}=(g+g_{12})/2$ for linearly-polarized light. 
We correspondingly define $n_\text{2 circ} = n_2$ and $n_\text{2 lin} = (n_2+n_{12})/2$.
Therefore, measuring photon-photon interactions in both (circularly and linearly polarized) configurations allows to determine the self- and cross-interaction parameters $g$ and $g_{12}$, as shown next.

\subsection{Measurement of self and cross interactions}
\label{Sec:int_measurements}

 \begin{figure}
     \includegraphics[width=0.95\columnwidth]{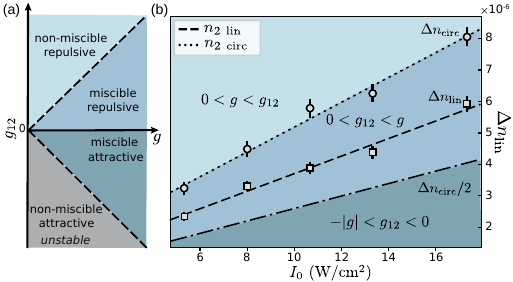}
     \caption{(a) Possible phases for a Bose mixture, as a function of $g_{12}$ and $g>0$.
     (b) Measured $\Delta n_\text{circ,lin}= -n_\text{2 circ,lin}I_0$ for a circularly (circles) and linearly (squares) polarized field, as a function of the input beam intensity $I_0$, for $\Delta=-2.2$~GHz and $T_\text{cell}$ = 165 °C. 
     The linear fits of $\Delta n_\text{circ}$  (dotted line) and $\Delta n_\text{lin}$  (dashed line) give $n_\text{2 circ}=-3.9 \pm 0.3 \times 10^{-7} $ cm²/W, $n_\text{2 lin}=-2.8 \pm 0.2 \times 10^{-7} $ cm²/W hence $g_{12}/g$ = 0.41 $\pm$ 0.14.
     With this diagram, it is possible to know the experimental phase of a mixture by looking at the position of the $\Delta n_\text{lin}$ dashed line with respect to the other lines: $\Delta n_\text{circ}$ and $\Delta n_\text{circ}/2$ (dotted-dashed line).
     The different colored regions show the different possible regimes as in figure (a).
    With the parameters of our experiment we observe $\Delta n_\text{circ}>\Delta n_\text{lin}>\Delta n_\text{circ}/2$, implying that the mixture is miscible with repulsive inter-component interactions.}
     \label{Fig:miscible}
\end{figure}

The physics of the two-component fluid is highly dependent on the sign and relative weight of $g$ and $g_{12}$. 
In our system, we work at negative detuning from the atomic transition, so that the circularly and linearly polarized fluids are stable (self-defocusing regime), giving both $g_\text{circ}=g>0$ and $g_\text{lin}=(g+g_{12})/2>0$. 
The phase diagram of a Bose mixture with $g>0$ is shown in Fig. \ref{Fig:miscible}(a): the fluid can either be in a miscible phase ($g>|g_{12}|$), where both components homogeneously occupy the entire available space, or in a non-miscible phase ($g<g_{12}$) where the two components are spatially separated.
In the miscible phase, we expect two branches of dispersion associated with the propagation of density and spin waves, whereas in the non-miscible phase the spin mode becomes unstable \cite{Pitaevskii2016, abad2013study}.


In this section, we first characterize the interaction parameters of \emph{single-component} fluids of light in linearly and circularly polarized states, to measure $g$ and $g_{12}$ and thus identify in which phase our system typically operates.
To this aim, we use the Bragg pulse spectroscopy method developed in \cite{piekarski2021measurement} to retrieve the dispersion of circularly and linearly polarized beams, given by the Bogoliubov expression:
\begin{equation}
    \Omega(k_p)=\sqrt{\Big(\dfrac{k_p^2}{2k_0}\Big)^2+ k_p^2 \Delta n_\text{circ,lin}}\ .
    \label{eq:Bogoliubov_optics}
\end{equation}
In this relation, $\Delta n_\text{circ,lin}=g_\text{circ,lin} |E_0|^2/k_0 = -n_\text{2 circ,lin}I_0$ is the non-linear refractive index of the vapor, with $I_0 = \epsilon_0 c |E_0|^2/2$ being the background intensity. 
The speed of sound of the superfluid is given by  $\smash{c=\sqrt{\Delta n_\text{circ,lin}}}$. 

Our setup consists of a 795 nm laser beam propagating through a  5~cm rubidium 87 vapor cell. 
We show in Fig.~\ref{Fig:miscible}(b) the measured values of $\Delta n_\text{circ}$ and $\Delta n_\text{lin}$ as a function of $I_0$, for a vapor cell temperature $T_\text{cell}$~=~165°C and a detuning $\Delta$~=~-2.2~GHz.
$\Delta$ is counted from the $F_g =  2$ to $F_e=2$ transition of the $D_1$ line of $^{87}$Rb.
Linear fits of the results give $n_\text{2 circ}=-3.9~\pm~0.3~\times~10^{-7} $~cm²/W and  $n_\text{2 lin}=-2.8~\pm~0.2~\times~10^{-7} $~cm²/W.
The measurable interaction difference in the circularly and linearly polarized fields is a first signature of the coupled fluid picture (\ref{eq:cnlse}).
We obtain $\Delta n_\text{circ}>\Delta n_\text{lin}>\Delta n_\text{circ}/2$, which is equivalent to $g>g_{12}>0$, showing that our fluid operates in the repulsive miscible regime where spin and density waves are observable.
We have checked that this remains the case for other values of the cell temperature and of the negative detuning.
This measurement also provides a first estimate of the ratio $g_{12}/g=(2\Delta n_\text{lin}-\Delta n_\text{circ})/\Delta n_\text{circ}\simeq 0.41 \pm 0.14$ for these experimental parameters.
This value will be used as a reference for measurements performed on a two-component fluid of light, that are presented in the next section.

\subsection{Observation of spin and density modes in a fluid of light mixture}
\label{sec:spin_density_modes_observation}

We now investigate the excitation and detection of density and spin modes in our miscible two-component fluid of light. 
We first study the decoupling between these two modes to demonstrate that they constitute eigenmodes of the system.
Following the paraxial fluid of light framework, we create a superfluid at the input of the non-linear medium ($z=0$) and excite it with a weak plane-wave perturbation propagating in the transverse plane on top of the background density.
Precisely, the expression of the optical field as the cell entrance is given in  the circular polarization basis $\{\sigma_+, \sigma_-\}$ by
\begin{align}
\begin{pmatrix}
E_+ \\
E-
\end{pmatrix}
= E_0
\begin{pmatrix}
-1/\sqrt{2} - \varepsilon\cos{\theta}e^{ik_px} \\
1/\sqrt{2} + \varepsilon\sin{\theta}e^{ik_px}
\end{pmatrix},
\label{eq:input_state}
\end{align}
with~$\varepsilon\ll1$.
The background field, of intensity $I_0 =\epsilon_0 c |E_0|^2/2$, is linearly polarized, forming a balanced mixture of the two components, while the polarization of the perturbation depends on the angle $\theta$.
The density and spin variables of the optical-fluid mixture are respectively defined as $I_\text{d} = \epsilon_0 c (|E_+|^2 + |E_-|^2)/2$ and $I_\text{s} = \epsilon_0 c(|E_-|^2 - |E_+|^2)/2$. 
The corresponding density and spin perturbations at the entrance of the  cell are
\begin{align}
    \delta I_\text{d}(x,y,z=0) &= 2\varepsilon I_0\sin{(\theta + \pi/4)}\cos{(k_px)}
    \label{initial_perturbation_d}\\
    \delta I_\text{s}(x,y,z=0) &= 2\varepsilon I_0\sin{(\theta -\pi/4)}\cos{(k_px)} \text{ ,}
    \label{initial_perturbation_s}
\end{align}
where the subscript $\text{d}$($\text{s}$) refers to the density (spin) mode.
The value of $\theta$ controls the relative weight of the initial density and spin perturbations, allowing the excitation of only one mode or both. 
For instance, when $\theta=\pi$/4---corresponding to a linearly polarized perturbation aligned with the background polarization---only the density mode is excited. 
In contrast, when $\theta$ = 3$\pi$/4, i.e. for a linearly polarized perturbation perpendicular to the background's polarization, only the spin mode is excited.
For intermediate values of $\theta$, the perturbation is elliptically polarized, and both modes are excited. \\

\begin{figure}
    \includegraphics[width=1 \columnwidth]{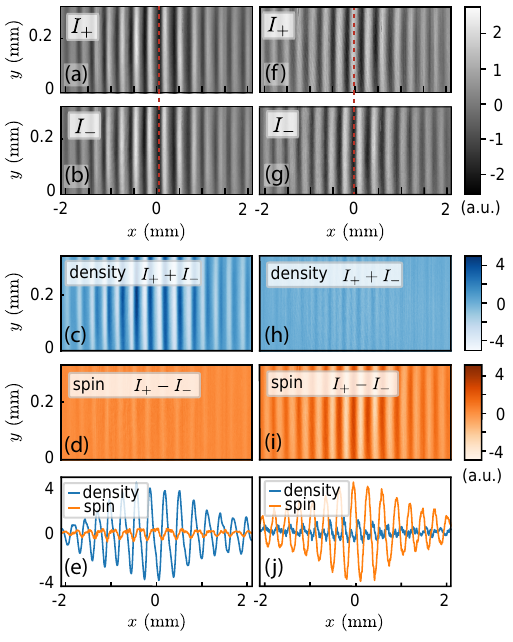}
    \caption{ (a,f) Output $\sigma_+$ ($I_+$) and (b,g) $\sigma_-$ ($I_+$) intensities (after subtraction of an non-modulated reference image) when imprinting a weak transverse perturbation of momentum $k_p= 21$~rad/mm on top of a linearly polarized background. (c,h) Resulting density ($I_+ + I_-$, in blue) and (d,i) spin ($I_+ - I_-$, in orange) images. 
    (e,j) shows the vertically integrated density and spin intensities.
    In (a-e) (left column), the input perturbation is linearly polarized, parallel to the background's polarization, so that only a density mode is injected. 
    In (f-j) (right column), the input perturbation is linearly polarized, perpendicular to the background's polarization, so that only a spin mode is injected.
    } 
    \label{Fig1_eigenmodes}
\end{figure}

Experimentally, we imprint the weak phase modulation with a spatial light modulator.
At cell input, the perturbation accounts for 5\% of the intensity of the background.
The beam's dimensions are 1.9 $\times$0.25~mm. 
It is elongated along the $x$-direction to limit horizontal defocusing, which would otherwise modify $k_p$ along propagation.
At cell output, the $\sigma_+$ and $\sigma_-$ components are split with a quarter waveplate and a polarizing beam splitter, and their intensities $I_+$ and $I_-$ are imaged on two different cameras.
The density and spin variables are measured by taking the sum and difference of the two images after subtracting an non-modulated reference image. 
The initial sinusoidal perturbation appears as a spatial modulation in both the intensities $I_+$ and $I_-$, associated with the $\sigma^+$ and $\sigma^-$  components of the light. 

In Figs. \ref{Fig1_eigenmodes}(a-e), we inject a pure density mode ($\theta = \pi/4$ in Eq. (\ref{eq:input_state})).
Figs. \ref{Fig1_eigenmodes}(a,b) show the corresponding output $I_+$ and $I_-$, for which the fringes appear in phase, and consequently are only visible in the density image (Fig. \ref{Fig1_eigenmodes}(c)) and not in the spin image (Fig. \ref{Fig1_eigenmodes}(d)).
Fig. \ref{Fig1_eigenmodes}(e) shows the vertical integration of those density and spin images, highlighting their difference in fringe visibility.
Conversely, Figs. \ref{Fig1_eigenmodes}(f-j) show the same observables when we inject a pure spin mode ($\theta = 3\pi/4$ in Eq. (\ref{eq:input_state})).
Here the fringes are out-of-phase in $I_+$ and $I_-$ Figs. (\ref{Fig1_eigenmodes}(f,g)) and thus appear only in the spin image (Fig. \ref{Fig1_eigenmodes}(i,j)) and not in the density image (Figs. \ref{Fig1_eigenmodes}(h,j))
We can thus selectively excite either mode and see that they remain decoupled during propagation within the cell.

This indicates that spin and density are indeed the eigenmodes of our system, which is a direct signature of the two-component behavior in our fluid of light
We now focus on investigating the dispersion spectrum of elementary excitations in this novel platform and compare it to that of atomic superfluids.

\subsection{Measuring the spin and density dispersions}
\label{sec:spin_density_modes}

\begin{figure*}
    \includegraphics[width=2\columnwidth]{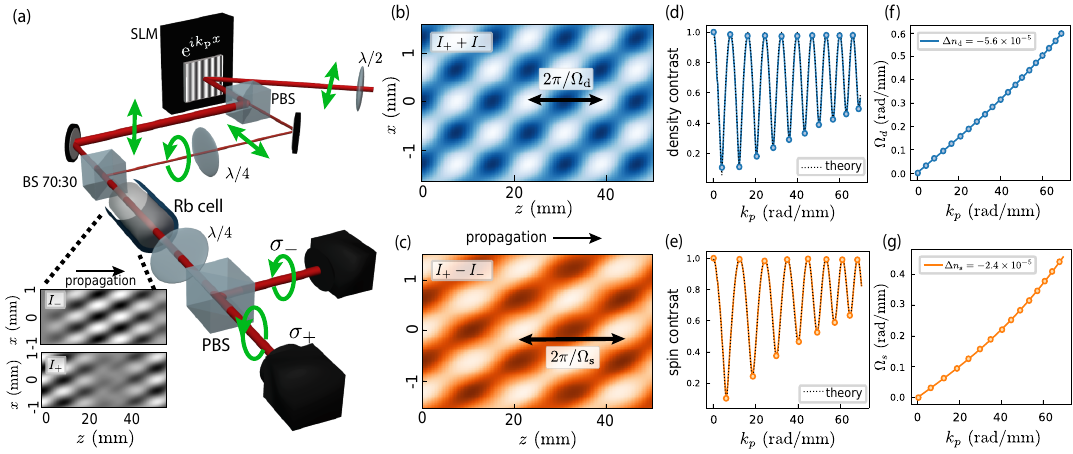}
    \caption{
    \label{Fig:setup}
    (a) Two-component Bragg spectroscopy setup.
   A circularly polarized perturbation carrying a transverse momentum $k_p$ is imprinted on top of a linearly polarized background with an SLM, and imaged at the cell input.
    At the cell output, the two circular  polarization components are split and the $I_+$ and $I_-$ intensities are imaged on two different cameras, of which we take the sum and the difference to obtain the density and spin intensities (for details, see Appendix \ref{Sec:appendix_setup}).
    The two intensity patterns in inset show $I_+$ and $I_-$ in the ($x$,$z$) plane \cite{Aladjidi2024_NLSE}, obtained from simulations of the coupled NLSEs for $k_p$ = 50 rad/mm. 
   The resulting (b) $I_\text{d}=I_++I_-$ and (c) $I_\text{s}=I_+-I_-$ in the ($x$,$z$) plane oscillate harmonically along $z$, with different frequencies respectively corresponding to $\Omega_\text{d}(k_p)$ and $\Omega_\text{s}(k_p)$.
    (d) Contrasts of the density and (e) spin fringes as a function of $k_p$ computed numerically at the cell output ($z=L$), of which we spot the extrema. The dotted black curves are the
      theoretical expectations (\ref{eq:fringes_amplitudes}). 
    (f) Reconstruction of the density and (g) spin dispersions from the extrema of contrast.
   Fitting data points with the Bogoliubov relation (\ref{eq:BG_mix}) (full blue in (f) and orange curves in (g)) gives the expected values of $\Delta n_\text{d,s}$, Eqs (\ref{eq:DeltaN_d}) and (\ref{eq:DeltaN_s})}.
\end{figure*}

Following the proposal of \cite{Martone2021}, we imprint a weak plane-wave perturbation  $e^{ik_p x}$ carrying a transverse momentum $k_p$ onto a uniform, linearly polarized background field $E_0$.
In particular, we use a circularly polarized perturbation ($\theta$ = $\pi$/2) such that both modes are excited with an equal initial amplitude.
The two eigenmodes are associated with two dispersion branches in the system, the spin and density branches, which both follow a Bogoliubov relation \cite{Pethick_Smith_2008}:
\begin{equation}
    \Omega_\text{d,s}(k_p)=\sqrt{\Big(\dfrac{k_p^2}{2k_0}\Big)^2+ k_p^2 \Delta n_\text{d,s}}\ \ ,
    \label{eq:BG_mix}
\end{equation}
with the `density' and `spin' non-linear refractive indices given by 
\begin{align}
    \Delta n_\text{d} &= \dfrac{g+g_{12}}{2k_0}|E_0|^2= \dfrac{n_2+n_{12}}{2}I_0,
    \label{eq:DeltaN_d}\\
    \Delta n_\text{s} &= \dfrac{g-g_{12}}{2k_0}|E_0|^2= \dfrac{n_2-n_{12}}{2}I_0 \text{ .}
    \label{eq:DeltaN_s}
\end{align}
The corresponding density and spin speeds of sound are  defined as
\begin{align}
c_\text{d,s} = \sqrt{\Delta n_\text{d,s}}.
\end{align}

To measure this two-branch dispersion, we extend the Bragg spectroscopy method to the two-component fluid.
The setup is shown in Fig. \ref{Fig:setup}(a) and is explained in more detail in the Appendix \ref{Sec:appendix_setup}.
At the entrance of the cell, an effective quench of the interaction (there are no interactions outside the cell at $z<0$, \cite{steinhauer2022analogue}) generates counter-propagating phonons at $-k_p$ that are stimulated by the input perturbation at $+k_p$, and result in the superposition of two counter-propagating plane waves with momenta $\pm k_p$ in the transverse plane $(x,y)$.
This effect leads to interference along the $z$ axis \cite{fontaine2020interferences} that we  numerically simulate  in Fig. \ref{Fig:setup}(b) \cite{Aladjidi2024_NLSE}.
This figure shows that  interferences along $z$ are non-harmonic for $I_+$ and $I_-$, but become harmonic for $I_+ + I_-$ (density) and $I_+ - I_-$ (spin), with frequencies corresponding to  $\Omega_\text{d}(k_p)$ and $\Omega_\text{s}(k_p)$, respectively.

Since we cannot image those interferences along $z$ (i.e. inside the cell), we use the output signal of the transverse fringes to retrieve the dispersion.
We scan $k_p$ and measure the contrast of the transverse fringes at the output of the cell, in both the intensity channels $I_\text{d}$ and $I_\text{s}$.
We show in Appendix \ref{Sec:appendix_theory} that these contrasts evolve as
\begin{align}
    \mathcal{C}_\text{d,s}(k_p,z)=\sqrt{2}\epsilon I_0
    \sqrt{1-\sin^2(\Omega_\text{d,s}(k_p)z)\Big[1\!-\!\frac{k_p^2/2k_0}{\Omega_\text{d,s}(k_p)}\Big]},
    \label{eq:fringes_amplitudes}
\end{align}
with $\mathcal{C}_d$ and $\mathcal{C}_s$ being  the contrasts of the transverse fringes for the density and spin channels.
This relation indicates that if, for a given $k_p$, an extremum of contrast is observed at the cell output $z=L$, the longitudinal frequency of the fringes is given by $\Omega_\text{d,s} = p \pi/2L$, where $p$ is an integer.
To retrieve the density and spin dispersion branches, we plot the density and spin contrast as a function of $k_p$ and spot the consecutive extrema, to which we attribute the dispersion value $\omega = p \pi / 2L $ for the $p$-th extremum. 
We fit the resulting plots with the Bogoliubov formula (\ref{eq:BG_mix}) and extract $\Delta n_\text{d}$ and $\Delta n_\text{s}$. 
The principle of the method is illustrated in Figs. \ref{Fig:setup}(d-g) through numerical simulations where we set $I_0 =1.6\times 10^5$~W/m$^2$, $n_2 = -5\times 10^{-6}$~cm$^2$/W  and $n_{12}=0.4n_2$. 
From the fringe contrasts of the density and spin signals in Fig.~\ref{Fig:setup}(d-e), we extract two distinct Bogoliubov branches shown in Fig. \ref{Fig:setup}(f-g). 
Their fit provides the expected values $\Delta n_\text{d}=5.6\times 10^{-5}=-(n_1+n_{12})I_0/2$ and $\Delta n_\text{s}=2.4\times 10^{-5}=-(n_1-n_{12})I_0/2$. 
The agreement with Eq.~(\ref{eq:fringes_amplitudes}) is also verified.

\begin{figure}
    \includegraphics[width=\columnwidth]{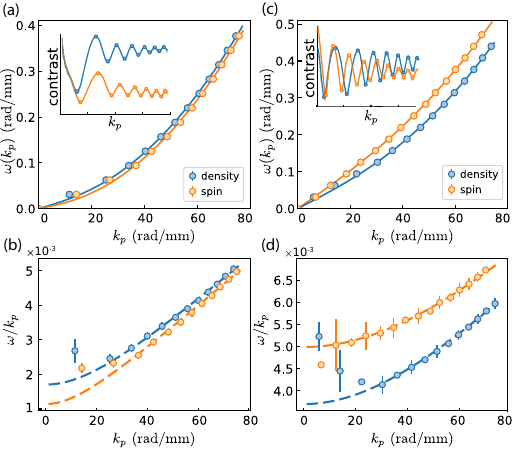}
    \caption{(a) Experimentally measured density (blue) and spin (orange) dispersion branches, at a detuning $\Delta = -1.8$\,GHz, an input beam power $P = 20$\,mW and $T_\text{cell}=165$\,°C. 
    The data points (blue and orange circles) are obtained from the extrema of  the density (blue) and spin (orange) fringe contrasts  shown in the inset. 
    The contrast extrema are spotted from the smoothed contrasts (dark blue and orange lines), behind which the raw data are visible in lighter colors.
    Solid dispersion curves are  fits to the Bogoliubov formula, which give  $\Delta n_\text{d}= 2.9 \pm 0.2 \times 10^{-6}$ and $\Delta n_\text{s} = 1.3 \pm 0.2 \times 10^{-6}$ \cite{footnote}.
    (b) Phase velocity $\Omega_\text{d,s}/k_p$ corresponding to (a). The circles show  at the extrema of contrast, the dashed lines show the theory corresponding to the fitted values of $\Delta n_\text{d,s}$.
    (c) Density and spin dispersions and (d) phase velocity for $\Delta = -2$\,GHz, $P = 900$\,mW and $T_\text{cell}=140$\,°C. In this case, we obtain  $\Delta n_\text{d}= 1.4 \pm 0.02 \times 10^{-5}$ and $\Delta n_\text{s} = 2.5 \pm 0.01 \times 10^{-5}$.
    }
    \label{Fig:disp_highI}
\end{figure}

Experimentally, the fringe contrast $\mathcal{C}_\text{d,s}$ is calculated from a sinusoidal fit over a central region covering four transverse wavelengths, divided by the unmodulated background.
Fig.~\ref{Fig:disp_highI}(a) shows the results for $\Delta=-1.8$~GHz, $T_\text{cell}$~=~165$^{\circ}$C and an input beam power of 20~mW. 
We extract the dispersion data points from the density and spin extrema of contrast and fit them with the Bogoliubov expression, which gives $\Delta n_\text{d}= 2.9 \pm 0.2 \times 10^{-6} $ and $\Delta n_\text{s} = 1.3 \pm 0.2 \times 10^{-6}$.
We obtain $c_\text{s}<c_\text{d}$, as expected from Sec.~\ref{Sec:int_measurements}, which showed that  $0<g_{12}<g$. 
In Fig.~\ref{Fig:disp_highI}(b), we plot the phase velocity $\omega/k_p$ for each modes, which clearly show the splitting of the two branches.

Surprisingly, we obtain significantly different results when performing the same measurement at high power,  specifically at high values of $I/I_\text{sat}$, where $I_\text{sat}$ is the saturation intensity of the medium at our detuning.
Figure \ref{Fig:disp_highI}(c) illustrates this, with spin and density dispersion branches measured for a higher input beam power of 900~mW, a lower vapor temperature of 140$^{\circ}$C (which decreases $I_\text{sat}$), and a detuning $\Delta=-2$~GHz.
Under these conditions, we reach a different regime where the spin branch is above the density branch, i.e. $c_\text{s}>c_\text{d}$, despite having measured $g_{12}>0$.
This phenomenon arises from beyond-two-body interactions in the coupled evolution equations of the two-component fluid, and is explored theoretically and experimentally in the next section.



\section{Controlling optical-fluid mixtures via saturation}
\label{sec:theory}

Having established our ability to observe a quantum fluid of light mixture and measure its characteristic two-branch dispersion, we now show how the saturation of the optical non-linearity in atomic vapor can be leveraged to precisely control the spectral properties of the mixture.
\subsection{Theory}
\label{Sec:theory}

To capture the complexity  and potential of our system, it is essential to take into account the saturation of the atomic vapor.
Saturation manifests itself as effects beyond the two-body contact photon-photon interaction as it encompasses higher order non-linear terms of the atomic susceptibility.

In the presence of saturation, the coupled NLSEs are modified as
\begin{align}
    i \partial_z E_\pm \!=\! \Big(\! -\frac{1}{2k_0} \nabla_\perp^2 +  \dfrac{g|E_\pm|^2+g_{12}|E_\mp|^2}{1+{I}/{I_\text{sat}}} \Big) E_\pm \text{ ,}
    \label{eq:cnlse_sat} 
\end{align}
where $I=\epsilon_0 c(|E_+|^2+|E_-|^2)/2$ is the total intensity, and $I_\text{sat}$ is the saturation intensity of the vapor.  
The dynamics of the fluid mixture is most conveniently described within a hydrodynamic formalism, which is based on the intensity-phase representation $E_\pm=2/(\epsilon_0 c)\sqrt{I_\pm}\exp(i\varphi_\pm)$ of the field components.
Inserting this representation into Eq. (\ref{eq:cnlse_sat}), we find a continuity equation for the intensity components $I_\pm$,
\begin{equation}
    \partial_z I_\pm+\nabla_\perp(I_\pm \boldsymbol{v}_\pm)=0,
    \label{eq:continuity}
\end{equation}
and an Euler equation for the velocity components $\boldsymbol{v}_\pm=(1/k_0){\nabla}_\perp\varphi_\pm$,
\begin{equation}
\partial_z\boldsymbol{v}_\pm
\!+\!\frac{\nabla_\perp \boldsymbol{v}_\pm^2}{2}\!=\!
\frac{\nabla_\perp}{k_0}
\Big(\frac{\nabla_\perp^2\sqrt{I_\pm}}{2k_0\sqrt{I_\pm}}
-\dfrac{\epsilon_0 c}{2}\frac{gI_\pm+g_{12}I_\mp}{1+I/I_\text{sat}}\Big).
\label{eq:Euler}
\end{equation}
Following the experiment, density and spin modes can be  probed by slightly perturbing the polarization components of a homogeneous fluid of light. 
To model this, we consider an unperturbed, linearly polarized fluid of total intensity $I_0$ ---see Eq. (\ref{eq:input_state})---, to which we add small intensity fluctuations $\delta I_\pm$ in the polarization components: $I_\pm=I_0/2+\delta I_\pm$. 
In the ensuing dynamics of the fluid of light,  spin and density modes  respectively correspond to fluctuations of $\delta I_\text{s}=\delta I_--\delta I_+$ and $\delta I_\text{d}=\delta I_++\delta I_-$. 
Assuming  $\delta I_\pm\ll I_0$, we  linearize Eqs. (\ref{eq:continuity}) and (\ref{eq:Euler}) and obtain the equations of motion for the spin and density modes. In Fourier space they read:
\begin{equation}
\label{eq:EOM}
    \partial^2_z\delta I_\text{d,s}(\boldsymbol{k},z)+\Omega_\text{d,s}^2(k)\delta I_\text{d,s}(\boldsymbol{k},z)=0,
\end{equation}
where the Bogoliubov dispersion $\Omega_\text{d,s}(k)$ is defined by Eq. (\ref{eq:BG_mix}), and the density and spin speeds of sound $c_\text{d,s}$ are now given by
\begin{align}
    c_{d}^2=\Delta n_\text{d} &= \dfrac{g+g_{12}}{2k_0}\dfrac{|E_0|^2}{(1+{I_0}/{I_\text{sat}})^2}
     \label{eq:sat_cd}\\
    c_{s}^2 =\Delta n_\text{s}&= \dfrac{g-g_{12}}{2k_0}\dfrac{|E_0|^2}{1+{I_0}/{I_\text{sat}}}.
    \label{eq:sat_cs}
\end{align}
Solutions of the wave equations (\ref{eq:EOM}) for the initial conditions (\ref{initial_perturbation_d}) and (\ref{initial_perturbation_s}) used in the experiment are detailed in Appendix \ref{Sec:appendix_theory} for clarity.
Compared to the weak non-linear regime explored in Sec. \ref{Sec:fluidmix_exp}, a key additional feature of the saturated fluid mixture is the different scaling of $c_\text{d}$ and $c_\text{s}$ with the background intensity $I_0/I_\text{sat}$. 
This implies that even when $g_{12}$ is positive, the spin dispersion branch can rise above the density dispersion branch if the background intensity exceeds  the threshold value $I_\text {thr} = 2I_\text{sat}/(g/g_{12}-1)$.

\subsection{Experiment}
\label{Sec:experiment_sat}
\begin{figure}
\includegraphics[width=\columnwidth]{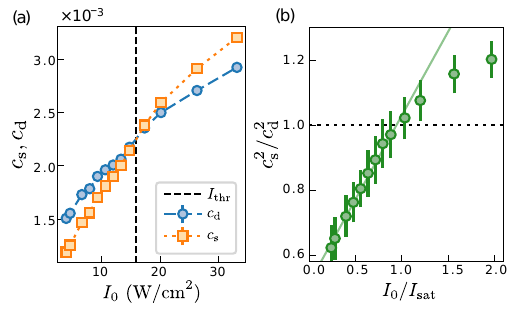}
\caption{
(a) Spin (blue circles) and density (orange squares) speeds of sound measured in our saturable optical-fluid mixture, as a function of the laser intensity. At low intensity $c_\text{s}<c_\text{d}$, while in the saturated regime $c_\text{s}>c_\text{d}$. The vertical dashed line indicates the theoretical crossing point calculated with the fitted values of $g_{12}/g$ and $I_\text{sat}$ obtained from (b).
(b) Ratio $(c_\text{s}/c_\text{d})^2$. For not too strong intensity we observe a linear increase consistent with Eq. (\ref{eq:cratio}). The intensity axis is normalized to $I_\text{sat}= 17 \pm 3$~W/cm$^2$, determined from the linear fit (solid green line) of $(c_\text{s}/c_\text{d})^2$ with intensity, which also gives $g_{12}/g=0.32 \pm 0.05$.  The data was taken for $\Delta = -2.2$~GHz and $T_\text{cell}$~=~165$^{\circ}$C. 
\label{Fig:cd_cs_crossover}}
\end{figure}
The speed-of-sound saturation phenomenon described by Eqs. (\ref{eq:sat_cd} and (\ref{eq:sat_cs}) is observed experimentally in Fig.~\ref{Fig:disp_highI}(c)-(d), where $\Delta n_\text{d}= 1.4 \pm 0.02 \times10^{-5} $ and $\Delta n_\text{d}= 2.5 \pm 0.01 \times10^{-5}$.
In practice, this provides a way to optically control the density and spin relative speed of sounds by adjusting $I_0/I_\text{sat}$, and to probe the transition where the spin branch rises above the density branch. 
This is the task that we now undertake experimentally.

To observe the transition from $c_\text{s}<c_\text{d}$ to $c_\text{s}>c_\text{d}$, we operate at a high cell temperature ($T_\text{cell}$=165$^{\circ}$C) and further detuned from resonance ($\Delta=-2.2$~GHz) to increase the effective $I_\text{sat}$. 
This allows us to achieve measurable interactions for low values of $I_0/I_\text{sat}$.
With these parameters, we have measured a set of values $c_\text{d}$ and $c_\text{s}$ for an input beam power ranging from 30~mW ($4$ W/cm$^2$) to 250~mW ($33$ W/cm$^2$ ), see Fig.~\ref{Fig:cd_cs_crossover}(a). 
At low intensity, we find that $c_\text{s}<c_\text{d}$, while in the saturated regime the spin and density branches cross and $c_\text{s}>c_\text{d}$, in full agreement with Eqs.~(\ref{eq:sat_cd}) and (\ref{eq:sat_cs}). 
Taking the ratio of these equations, we infer:
\begin{equation}
\label{eq:cratio}
    \frac{c_\text{s}^2}{c_\text{d}^2}=\dfrac{1-g_{12}/g}{1+g_{12}/g}\Big(1+\frac{I_0}{I_\text{sat}}\Big),
\end{equation}
which predicts a linear scaling of the squared ratio of speeds of sound with the laser intensity.
This is indeed what we observe in Fig.~\ref{Fig:cd_cs_crossover}(b). 
We attribute the deviation from linear scaling at high intensity to the phenomenon of self defocusing in the vertical $y$ direction, which effectively decreases the background intensity felt by the spin and density waves.  
Following Eq.~(\ref{eq:cratio}), we extract both the saturation intensity $I_\text{sat} =  17 \pm 3 $ W/cm$^2$ and $g_{12}/g = 0.32 \pm 0.05$ from a linear fit of the experimental ratio $(c_\text{s}/c_\text{d})^2$ as a function of $I_0$.
From these fitted values, we calculate the threshold intensity $I_\text{thr}=2I_\text{sat}/(g/g_{12}-1)$ where the two dispersion branches are expected to cross. 
The obtained value of $I_\text{thr}=15.8$~W/cm$^2$ is indicated by the vertical dotted line in Fig. \ref{Fig:cd_cs_crossover}(a) and accurately pinpoints the crossing point.
This fully corroborates the theoretical interpretation of the role of saturation, and demonstrates experimental ability to optically control the relative energy of the density and the spin modes, independently of the interaction strength.




\section{Summary and conclusion}
\label{Sec:summary}

Our work demonstrates the first experimental realization of a two-component fluid of light.
We found clear signatures of the coupled nonlinear evolution of the two circular polarization components, showing that they follow the coupled Gross-Pitaevskii equations.
Through selective excitation and detection of the density and spin modes, we confirmed their nature as eigenmodes of the system, as expected for a miscible Bose-Bose mixture.
Using a Bragg-like spectroscopy technique, we simultaneously measured the density and spin dispersion branches, along with the associated speeds of sound.
We also discovered a novel feature compared to atomic two-component Bose-Einstein condensates: the inversion of the density and spin dispersion branches, caused by the saturation of the nonlinear medium.
This saturation, microscopically due to beyond-two-body contact interaction terms, enabled us to tune the relative density and spin spectral response of the system.
Specifically, we found that in the highly saturated regime where the spin speed of sound exceeds the density speed of sound, the system reproduces the phenomenology of a miscible mixture with attractive inter-component interactions. 
These results open up many avenues for studying the physics of binary Bose mixtures in different regimes.

Future investigations will focus on the study of superfluidity in two-component systems.
Spin superfluidity is for instance attracting significant interest, particularly as a way to observe the Andreev-Bashkin effect, a beyond-mean-field phenomenon describing  non-dissipative drag between two coupled superfluid flows \cite{Karle_coupled_superfluidity, Nespolo_2017, Utesov_2018,sekino_2022_spinConductivity,sekino_2023,carlini_2021_SpinDrag,fava_2018_spinSF}.
Exploring regimes where only the density or spin mode is superfluid could also offer advantages for analog gravity experiments \cite{berti2022superradiantphononicemissionanalog}.

By exploiting other resonance lines and interaction regimes of the rubidium vapor, one could access the non-miscible phase of the two-component fluid.
This would enable exploration of the miscible to non-miscible phase transition, including the spontaneous formation of domains \cite{Spielman2014} and their associated universal scaling laws \cite{Hofmann_2014, ji2023domainformationuniversallycritical}.
In the non-miscible regime, one could study the formation and stability properties of massive vortices \cite{Richaud_2020, Patrick_2023}, as well as quantum turbulence \cite{silva_2021_tur2comp} and hydrodynamic instabilities \cite{Sasaki_2009_RayleighTaylor} unique to this system.

\section{Acknowledgments}

The authors acknowledge insightful discussions with Giovanni Martone, Iacopo Carusotto, Nicolas Pavloff, Pierre-Elie Larré and Alberto Bramati. We thank financial support from the Agence Nationale de la Recherche (NC for grant ANR-19-CE30-0028-01 CONFOCAL and QG for grant ANR-21-CE47-0009 Quantum-SOPHA). QG is member of the Institut Universitaire de France (IUF).

\appendix
\section{Amplitude of spin and density waves}
\label{Sec:appendix_theory}

In this appendix, we describe the evolution of spin and density waves in an optical fluid mixture with the initial condition (\ref{eq:input_state}), and derive exact expressions for their contrast $\mathcal{C}_\text{d,s}$, used in the experiment to extract Bogoliubov dispersions.

The spin and density waves are  solutions of the equations of motion (\ref{eq:EOM}). For the sake of simplicity, from now on we assume a circularly polarized perturbation, $\theta=\pi/2$, which corresponds to exciting the spin and density modes with equal initial amplitudes. Solving Eq. (\ref{eq:EOM}) 
 requires four initial conditions, the values of $\delta I_\text{d,s}$ and $\partial_z\delta I_\text{d,s}$ at $z=0$. From Eqs. (\ref{initial_perturbation_d}) and (\ref{initial_perturbation_s}), the initial value of $\delta I_\text{d,s}$ follows as
 \begin{equation}
     \delta I_\text{d,s}(x,y,z=0)=\sqrt{2}\varepsilon I_0\cos(k_p x).
 \end{equation}
The initial condition for the $\partial_z\delta I_\text{d,s}$, on the other hand, is deduced from the initial phases $\varphi_-\pm\varphi_+=\sqrt{2}\varepsilon \sin(k_p x)$ by virtue of the relation
\begin{equation}
\label{eq:inicond}
    \delta I_\text{d,s}=-\frac{I_0}{2k_0}\nabla_\perp^2(\varphi_-\pm\varphi_+),
\end{equation}
which follows from the linearization of the continuity equation (\ref{eq:continuity}). This leads to
\begin{equation}
\label{eq:inicond_dz}
    \partial_z\delta I_\text{d,s}(x,y,z=0)=\frac{\varepsilon I_0 k_p^2}{2k_0}\sin(k_p x).
\end{equation}
Solving Eq. (\ref{eq:EOM}) together with the initial conditions (\ref{eq:inicond}) and (\ref{eq:inicond_dz}), we obtain the expression of the optical  spin and density waves at any position in the vapor. They take the form $ \delta I_\text{d,s}=\mathcal{C}_\text{d,s}(k_p,z)\cos[k_p x-\phi_\text{d,s}(k_p,z)]$, with a mode amplitude given by Eq. (\ref{eq:fringes_amplitudes}) of the main text.

\section{Setup details}
\label{Sec:appendix_setup}
Here we describe the setup of the two-component Bragg pulse spectroscopy  in more details.
The polarization of the beam that is sent onto the SLM is set to be linear at a nearly vertical angle so that only a small part of the beam ---the horizontally polarized component--- picks up the modulation, while the majority vertical component is simply reflected. 
Those two linear components are then split by a polarizing beam splitter, and the polarization of the perturbation beam is adjusted with a half- or quarter- waveplate.
The two beams are then recombined on a 30:70 beam-splitter in a Mach-Zehnder configuration. 
This configuration ensures a homogeneous phase difference between the background and the perturbation.
In total, the perturbation accounts for 5\% of the background intensity. 
The SLM screen is imaged at the cell input, where the perturbation appears as fringes along $x$, of periodicity $2\pi/k_p$.
The beam is elongated along the $x$-axis ($w_{0x}$=1.9 mm, $w_{0y}$=0.25 mm), in order to have a flat intensity along the $x$-axis, minimizing self-defocusing along this axis, as otherwise $k_p$ would be modified along propagation.
At the cell output, the two circular polarization components are split by a quarter-waveplate at an angle of 45° and a PBS, and imaged with a magnification of two on two identical cameras. 
\raggedright
\bibliography{bib2compDisp}

\begin{thebibliography}{65}
\bibitem [{foo()}]{footnote}%
  \BibitemOpen
  \href@noop {} {}\bibinfo {note} {At low beam power, Fig.~4(a), the visible vertical shift between the density and spin contrasts is due to absorption in the vapor, which at low intensity affects more significantly the spin contrast. At higher beam power, Fig.~4(c), the absorption saturates so the shift between the the density and spin contrasts is reduced.}\BibitemShut {Stop}%
\end{thebibliography}

\end{document}